\documentclass[superscriptaddress,twocolumn,
longbibliography,pra,aps,10pt]{revtex4-2}
\pdfoutput=1
\usepackage{graphicx}
\usepackage{amsmath}
\usepackage{amsfonts}
\usepackage{amssymb}
\usepackage{xcolor}
\usepackage{hyperref}
\hypersetup{colorlinks=true,allcolors=blue}
\usepackage[OT4]{fontenc}

\usepackage{footnote}
\usepackage{textcomp}

\usepackage{upgreek}
\usepackage{bm}

\DeclareUnicodeCharacter{2009}{\,}

\renewcommand{\t}[1]{\textrm{#1}}
\newcommand{\nn}{\nonumber\\}

\newcommand{\q}{\bm{q}}

\newcommand{\g}{\gamma}

\newcommand{\w}{\omega}

\newcommand{\D}{\Delta}

\renewcommand{\L}{\mathcal{L}}

\newcommand{\+}{^\dagger}
\renewcommand{\>}{\rangle}
\newcommand{\<}{\langle}

\newcommand{\G}{\vert G\>}
\newcommand{\X}{\vert X\>}

\newcommand{\XX}{\vert X\>\<X\vert}

\newcommand{\XG}{\vert X\>\<G\vert}
\newcommand{\GX}{\vert G\>\<X\vert}

\begin{document}

\title{$N$-photon bundle statistics in different solid-state platforms}

\author{M. Cosacchi}
\affiliation{Theoretische Physik III, Universit{\"a}t Bayreuth, 95440 Bayreuth, Germany}
\author{A. Mielnik-Pyszczorski}
\affiliation{Theoretische Physik III, Universit{\"a}t Bayreuth, 95440 Bayreuth, Germany}
\affiliation{Department of Theoretical Physics, Wroc{\l}aw 
University of Science and Technology, 50-370 Wroc{\l}aw, Poland}
\author{T. Seidelmann}
\affiliation{Theoretische Physik III, Universit{\"a}t Bayreuth, 95440 Bayreuth, Germany}
\author{M. Cygorek}
\affiliation{Heriot-Watt University, Edinburgh EH14 4AS, United Kingdom}
\author{A. Vagov}
\affiliation{Theoretische Physik III, Universit{\"a}t Bayreuth, 95440 Bayreuth, Germany}
\affiliation{ITMO University, St. Petersburg, 197101, Russia}
\author{D. E. Reiter}
\affiliation{Institut f\"ur Festk\"orpertheorie, Universit\"at M\"unster, 48149 M\"unster, Germany}
\author{V. M. Axt}
\affiliation{Theoretische Physik III, Universit{\"a}t Bayreuth, 95440 Bayreuth, Germany}

\begin{abstract}
The term $N$-photon bundles has been coined for a specific type of photon emission, where light quanta are released from a cavity only in groups of $N$ particles.
This emission leaves a characteristic number distribution of the cavity photons that may be taken as one of their fingerprints.
We study this characteristic $N$-photon bundle statistics considering two solid-state cavity quantum electrodynamics (cQED) systems.
As one example, we consider a semiconductor quantum-dot--microcavity system coupled to longitudinal acoustic phonons.
There, we find the environmental influence to be detrimental to the bundle statistics.
The other example is a superconducting qubit inside a microwave resonator.
In these systems, pure dephasing is not important and an experimentally feasible parameter regime is found, where the bundle statistics prevails.
\end{abstract}
\maketitle

\section{Introduction}
\label{sec:Introduction}

\begin{figure}[t]
	\centering
	\includegraphics[width=\columnwidth]{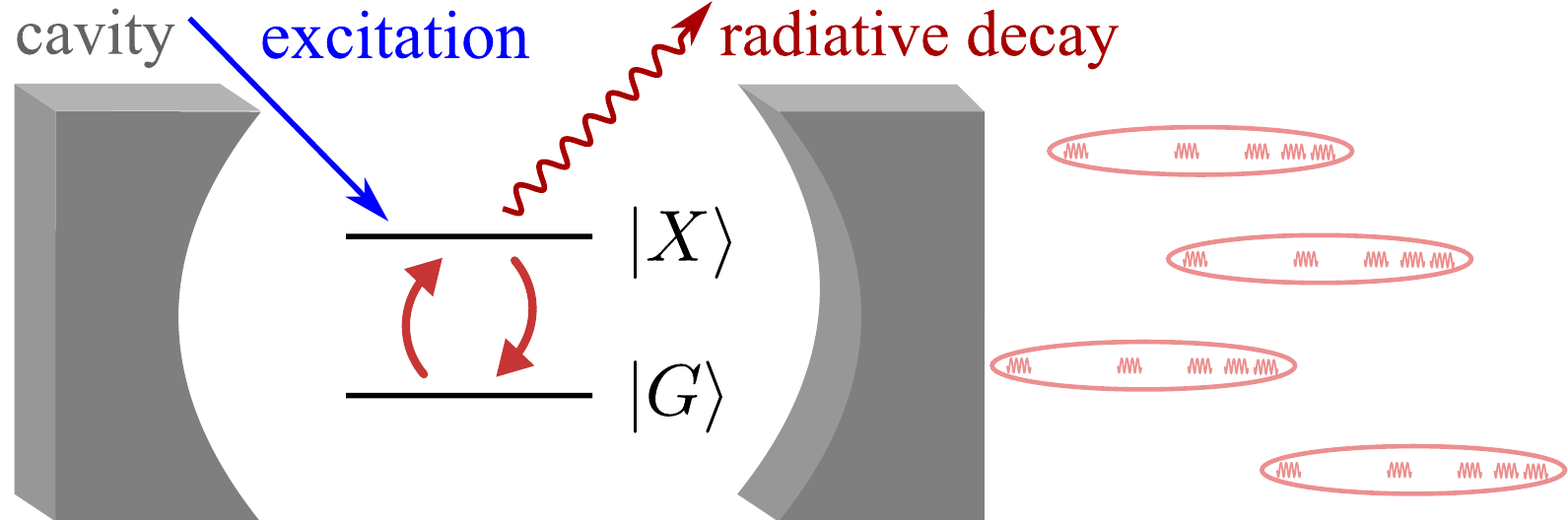}
\caption{Sketch of a two-level system (2LS) embedded in a cavity resulting in the coupling to one cavity mode.
The 2LS is driven by a continuous external excitation.
It can decay radiatively, while the cavity is lossy.
For particular sets of parameters, $N$-photon bundles leave the cavity.
They are characterized by the specific temporal spacing between the constituent photons and their specific photon number statistics.
Exemplary, four $5$-photon bundles are depicted.}
	\label{fig:sketch}
\end{figure}

Many innovative applications of the quantum realm rely on the on-demand preparation of specific, highly nonclassical target states.
Cavity quantum electrodynamics (cQED) is a machinery well suited for this purpose.
On numerous different platforms, e.g., atoms in resonators \cite{Cummings1989,Varcoe2000,Zhou2012}, superconducting qubits in microwave resonators \cite{Hofheinz2008,Hofheinz2009}, or semiconductor quantum dots in microcavities \cite{Michler2000,Santori2001,Santori2002,He2013,
Wei2014Det,Ding2016,Somaschi2016,
Schweickert2018,Hanschke2018,Cosacchi2019,
Akopian2006,Stevenson2006a,Hafenbrak2007,Dousse2010,
delvalle2013dis,Mueller2014,Orieux2017,Seidelmann2019,
Cosacchi2020b,Gea-Banacloche1990,Cosacchi2021a}, preparation of single photons, entangled photon pairs, Fock states, and Schrödinger or Voodoo cat states has been proposed or achieved.
Recently, a new class of emitters has been proposed \cite{Munoz2014,Bin2021}, where the photon emission takes place only in groups of an integer number $N$.
The term $N$-photon bundle has been coined to describe these multiphoton structures.
There are numerous ways to characterize these structures, e.g., in terms of their emission properties \cite{Munoz2018,Diaz-Camacho2021} or their internal correlations between the constituent photons \cite{LopezCarreno2018}, which can be interpreted as a consequence of their specific temporal spacing, see sketch in Fig.~\ref{fig:sketch}.
In contrast to the ordinary Fock state $|N\>$, a bundle is emitted as a cascade over successive Fock states $|n\>$, where $0\leq n\leq N$, which is a direct result of the outcoupling via resonator losses.
In a resonator with loss rate $\kappa$, the Fock state $|n\>$ effectively decays with the rate $n\kappa$, explaining the temporal spacing between the photons constituting the bundle.
As one feature of the $N$-photon bundle, we can find a characteristic photon statistics of the resonator photons in the ideal case of an $N$-photon bundle \cite{Munoz2014}
\begin{align}
\label{eq:1/n}
P_N(n)=
\begin{cases}
1-\frac{\<n\>}{N}\sum\limits_{j=1}^N\frac{1}{j} \quad& n=0\\
\frac{\<n\>}{N}\frac{1}{n} \quad& 1\leq n\leq N\\
0 \quad& n>N
\end{cases}
\end{align}
with $\<n\>$ being the average photon number in the resonator.
From a detection point of view this means the following:
When a Poissonian source emits photons, their arrival times at the detector are distributed randomly;
in the case of an $N$-photon bundle emission, the bundles arrive randomly, but the photons contained in each bundle obey the temporal order as sketched in Fig.~\ref{fig:sketch}.
Therefore, there is a Poissonian distribution over bundles.
In this sense, $N$-photon bundles can be considered as an alternative to Fock states as building blocks for more complex quantum states of light.
Furthermore, $N$-photon bundles have the property to herald a Fock state.
Finally, on timescales longer than the size of the bundle, Planck's constant is effectively renormalized in the relationship between frequency and energy, $E=N\hbar\w$.
Therefore, $N$-photon bundles are even discussed for medical applications due to a greater penetration depth and increased resolution \cite{Munoz2014}.
Even bundle generation using phonons instead of photons has been proposed \cite{Bin2020}.

In this work, we consider the bundle statistics in Eq.~\eqref{eq:1/n} as one of the possible ways to characterize a bundle and study this fingerprint in two different solid-state platforms:
(i) semiconductor quantum dots (QDs) in microcavities and (ii) superconducting qubits in microwave resonators.

In QDs, the coupling to longitudinal acoustic phonons is known as the main source of decoherence.
We therefore analyze a QD--cavity system coupled to a phonon environment modeled in a microscopic picture.
This full many-body problem is solved in a numerically exact way by employing a path-integral formalism.
We compare these results with those found in a model accounting for phonons only via a phenomenological pure dephasing rate.
For realistic parameters that are currently achievable, we find that the phonon influence leads to photon number distributions that deviate significantly from the bundle statistics in Eq.~\eqref{eq:1/n}.

In superconducting qubit--microwave resonator systems, pure dephasing is negligible.
For these systems, we propose a set of parameters experimentally well within reach, where the bundle statistics with $N=2$ is preserved.
We show that for this purpose a resonator with a mediocre $Q$-factor is optimal.

\section{Model and methods}
\label{sec:model}

\subsection{cQED model}
\label{subsec:cQED}

Both example systems can be described by a strongly driven Jaynes--Cummings model with the Hamiltonian in a frame co-rotating with the frequency of the external excitation $\w_{\t{L}}$ in the usual dipole and rotating-wave approximations
\begin{align}
H=\,&-\hbar\D\w_{\t{LX}}\XX
+\hbar\D\w_{\t{CL}} a\+ a\nn
&+\hbar g\left(\XG a + \GX a\+\right)\nn
&+\hbar f\left(\XG + \GX\right)\, .
\end{align}
The two-level system (2LS) has an excited state $\X$ at energy $\hbar\w_{\t{X}}$ and a ground state $\G$ at energy zero.
$a$ ($a\+$) is the annihilation (creation) operator of a photon in the single resonator mode at energy $\hbar\w_{\t{C}}$ coupled to the 2LS by $g$.
The detuning between the external excitation with strength $f$ and the upper state $\X$ is denoted by $\D\w_{\t{LX}}=\w_{\t{L}}-\w_{\t{X}}$ and the detuning between resonator and external excitation $\D\w_{\t{CL}}=\w_{\t{C}}-\w_{\t{L}}$ is defined analogously.
The detuning between resonator and the upper state $\X$, $\D\w_{\t{CX}}=\w_{\t{C}}-\w_{\t{X}}$, is fixed by the growth process of the structure.
Hence, we keep it constant in our analysis.

When the 2LS is strongly driven ($f\gg g$) and it is in the dispersive regime ($\D\w_{\t{CX}}\gg g$), a sharp $N$-photon resonance emerges with $N$ being an integer.
It corresponds to a polariton of the type $(|G,0\>\pm|X,N\>)/\sqrt{2}$, where $|\chi,n\>$ denotes the product state of the 2LS state $|\chi\>$ and the photon number state $|n\>$.
When dissipative channels are included by introducing the excited state's radiative decay with rate $\gamma$ and resonator losses with rate $\kappa$, this resonance becomes a source of $N$-photon bundles, when the stationary state is reached \cite{Munoz2014}.

We include these dissipative effects by accounting for the Lindblad superoperators $\L_{\vert G\rangle\!\langle X\vert,\gamma}$ and $\L_{a,\kappa}$ acting on the density matrix $\rho$ as
\begin{align}
\mathcal{L}_{O,\Gamma}\rho=\Gamma\left(O\rho O\+ -\frac{1}{2}\left\lbrace\rho,O\+ O\right\rbrace_+\right)\, ,
\end{align}
describing loss processes with rate $\Gamma$ on a dissipation channel $O$, where $\left\lbrace A,B\right\rbrace_+$ is the anti-commutator of operators $A$ and $B$.

\subsubsection{QD model}
\label{subsec:phonons}

At first, we consider a self-assembled GaAs QD system in a single-mode microcavity.
In these systems, additionally the pure-dephasing coupling of the electronic states to an environment of longitudinal acoustic phonons is important \cite{Reiter2014,Reiter2019}.
It is described by the Hamiltonian \cite{Besombes2001,Borri2001,
Krummheuer2002,Axt2005}
\begin{align}
\label{eq:H_Ph}
H_{\t{Ph}}=&\,\hbar\sum_{\q} \w_{\q} b_{\q}\+ b_{\q}\nn
&+\hbar\sum_{\q} \left(\g_{\q}^{\t{X}}b_{\q}\+ +\g_{\q}^{\t{X}*}b_{\q}\right)\XX\, ,
\end{align}
where $b_{\q}$ ($b_{\q}\+$) annihilates (creates) a phonon of energy $\hbar\w_{\q}$ in mode $\q$ with the coupling strength $\g_{\q}^{\t{X}}$.
The phonons are assumed to be initially in thermal equilibrium at temperature $T$.

This coupling to phonons is the source of many well-known effects in QDs, like the phonon sideband in the QD emission spectrum \cite{Besombes2001,McCutcheon2016}, the renormalization of the Rabi frequency \cite{Kruegel2005,Ramsay2010b}, and the damping of Rabi oscillations \cite{Foerstner2003,Machnikowski2004,Ramsay2010a}.
To discuss resonances, the most important effect is the polaron shift of the excited state $\X$.
Whenever we refer to the excited state energy when phonons are taken into account, we mean the polaron-shifted excited state energy.

To treat this full many-body Hamiltonian in a numerically exact way, we employ an iterative real-time path-integral formalism \cite{Makri1995a,Makri1995b} to solve the Liouville--von Neumann equation (details are explained in Refs.~\cite{Vagov2011,Barth2016,Cygorek2017}).
Within this approach, all effects mentioned above are thus taken into account.

Unless noted otherwise, we take $\hbar g=0.02\,$meV \cite{Najer2019}, $\gamma=1\,$ns$^{-1}$, and $\kappa=8.5\,$ns$^{-1}$ \cite{Schneider2016}.
These values, in particular, the cavity loss rate $\kappa$ are realistically achievable \cite{Schneider2016}.
The record in cavity quality so far is around $\kappa\approx4\,$ns$^{-1}$ to $6\,$ns$^{-1}$ \cite{Najer2019}, which means that it should be possible to achieve the value of $\kappa$ chosen here with current state-of-the-art equipment with reasonable effort.
Further, following Ref.~\cite{Munoz2014}, we set $\hbar\D\w_{\t{CX}}=-60\hbar g=-1.2\,$meV and $\hbar f=32\hbar g=0.64\,$meV.
For the phonon coupling, standard GaAs parameters \cite{Krummheuer2005,Cygorek2017} are chosen for a QD with a radius of $3\,$nm.

\subsubsection{Superconducting qubit model}
\label{subsec:SC}

As a second example, we consider a superconducting qubit in a microwave resonator.
Here, pure dephasing is negligible.
Therefore, no addition to the model in Sec.~\ref{subsec:cQED} is necessary.

We use the parameter set $\hbar g=0.079\,\upmu$eV, $\gamma=1.54\,\upmu$s$^{-1}$, $\kappa=0.29\,\upmu$s$^{-1}$, i.e. $\kappa\ll\gamma$, taken from Ref.~\cite{Hofheinz2009}.
Again, following Ref.~\cite{Munoz2014}, we choose $\hbar\D\w_{\t{CX}}=-60\hbar g=-4.71\,\upmu$eV and $\hbar f=32\hbar g=2.51\,\upmu$eV.


\section{Results: QD--cavity system}
\label{sec:QDCs}

\begin{figure}[t!]
	\centering
	\includegraphics[width=\columnwidth]{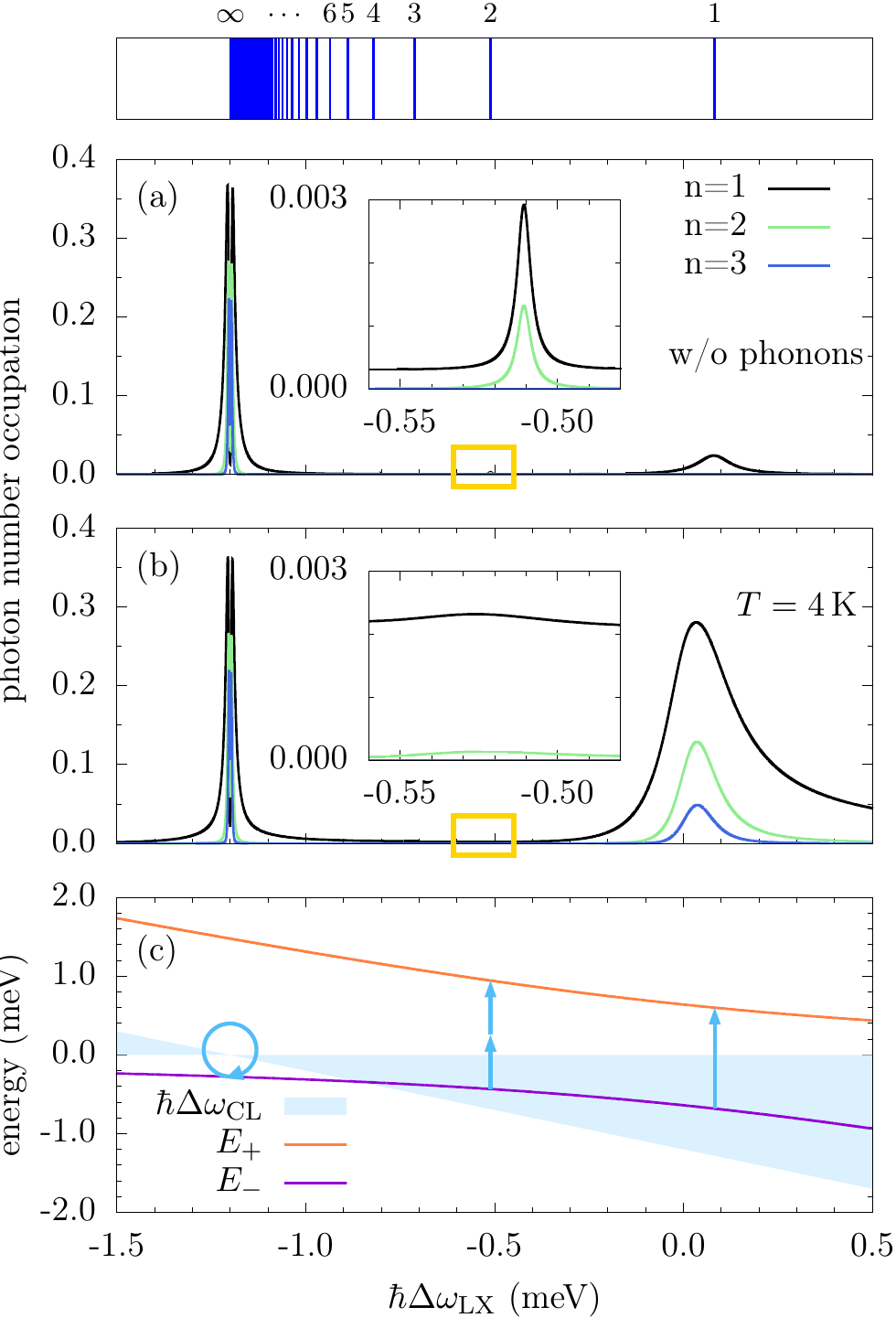}
\caption{Stationary photon number occupation in the QD--cavity system as a function of the laser--exciton detuning $\D\w_{\t{LX}}$ (a) without taking phonon effects into account, (b) including phonons initially at $T=4\,$K (the insets show the region marked by yellow boxes on a larger scale), (c) the corresponding energies of the laser-dressed states $|+\>$ and $|-\>$.
The energy of a photon  in the rotating frame is given by the cavity--laser detuning $\hbar\D\w_{\t{CL}}$, which is plotted as a shaded area to illustrate its modulus.
Arrows indicate the number of photons involved in the processes leading to the various resonance peaks, while their length corresponds to their energy $\hbar\D\w_{\t{CL}}$.
The circular arrow indicates a $1$-photon process with a photon energy (in the rotating frame) of $\hbar\D\w_{\t{CL}}=0$.
The blue lines above panel (a) mark the energetic positions of the bundle resonances, starting for $N=1$ and quickly converging to $\hbar\D\w_{\t{CX}}$ for larger $N$.
Since the bundle resonance is derived from the condition that $N$ cavity photons energetically fit between the two dressed states, an equation analogous to Eq.~\eqref{eq:bundle_resonance} can be found for the trivial case $N=1$.
}
	\label{fig:stat_occ_QDC_triple}
\end{figure}

\subsection{Resonance landscape and N=2}
\label{subsec:Resonances}

The resonance corresponding to an $N$-photon bundle is found at \cite{Chough2000,Munoz2014}
\begin{align}
\label{eq:bundle_resonance}
\D\w_{\t{LX}}=&\,\frac{\sqrt{4\left(N^2-1\right)f^2+N^2\D\w_{\t{CX}}^2}+\D\w_{\t{CX}}}{N^2-1}\nn
&+\D\w_{\t{CX}}\, .
\end{align}

In this work, we focus mostly on the case $N=2$.
For the QD--cavity system, this results in a detuning value of $\hbar\Delta\w_{\t{LX}}=-0.51\,$meV.
Higher-order bundles with $N>2$ can be reached by tuning the excitation to the corresponding resonance according to Eq.~\eqref{eq:bundle_resonance}, however for the realistic set of parameters assumed here they are negligible.

To illustrate the appearing resonances, we scan the stationary photon number occupation with the laser frequency $\w_{\t{L}}$.
Figure~\ref{fig:stat_occ_QDC_triple} shows the corresponding results for the photon numbers $n=1$, $2$, and $3$ in the QD--cavity system.
Three resonance peaks emerge in the vicinity of the bundle resonance [presented in Fig.~\ref{fig:stat_occ_QDC_triple}(a)], which itself is shown on a magnified scale in the inset.

The most prominent peaks are found for the limiting cases $N\to\infty$ and $N=1$.
For $N\to\infty$ a double-peaked structure emerges at $\hbar\D\w_{\t{LX}}=\hbar\D\w_{\t{CX}}=-1.2\,$meV (cf., Fig~\ref{fig:magnified_N_infty} for a zoom-in).
At its center the photon statistics is Poissonian and is hardly influenced by phonons [cf., Figs.~\ref{fig:stat_occ_QDC_triple}(a) and (b)].
In contrast, the peak at $\hbar\D\w_{\t{LX}}\approx0.08\,$meV corresponds to the resonance for $N=1$.
Here, Fock states with $n>1$ are not occupied due to a photon blockade effect [cf., Fig.~\ref{fig:stat_occ_QDC_triple}(a)], which is spoiled once phonons are considered:
then, the system can climb up the Jaynes-Cummings ladder [cf., Fig.~\ref{fig:stat_occ_QDC_triple}(b)].
The different physical mechanisms giving rise to these two limiting cases and the phonon influence on them is discussed in detail in Appendix~\ref{app:limits}.

We now consider the range of bundle physics for $1<N<\infty$ and focus on $N=2$.
The characteristic bundle statistics as denoted in Eq.~\eqref{eq:1/n} is well visible for the $2$-photon bundle shown in the inset of Fig.~\ref{fig:stat_occ_QDC_triple}(a), in particular, the $3$-photon occupation is zero.

To understand all the resonances, we diagonalize the Hamiltonian of the laser-driven 2LS neglecting the cavity (since $f\gg g$).
As a result, we obtain the dressed states $|+\>$ and $|-\>$.
Their energies in the laser-rotating frame are plotted in Fig.~\ref{fig:stat_occ_QDC_triple}(c) along with the energy of a cavity photon given by $\hbar\D\w_{\t{CL}}$ in this frame.


\begin{figure}[t]
	\centering
	\includegraphics[width=\columnwidth]{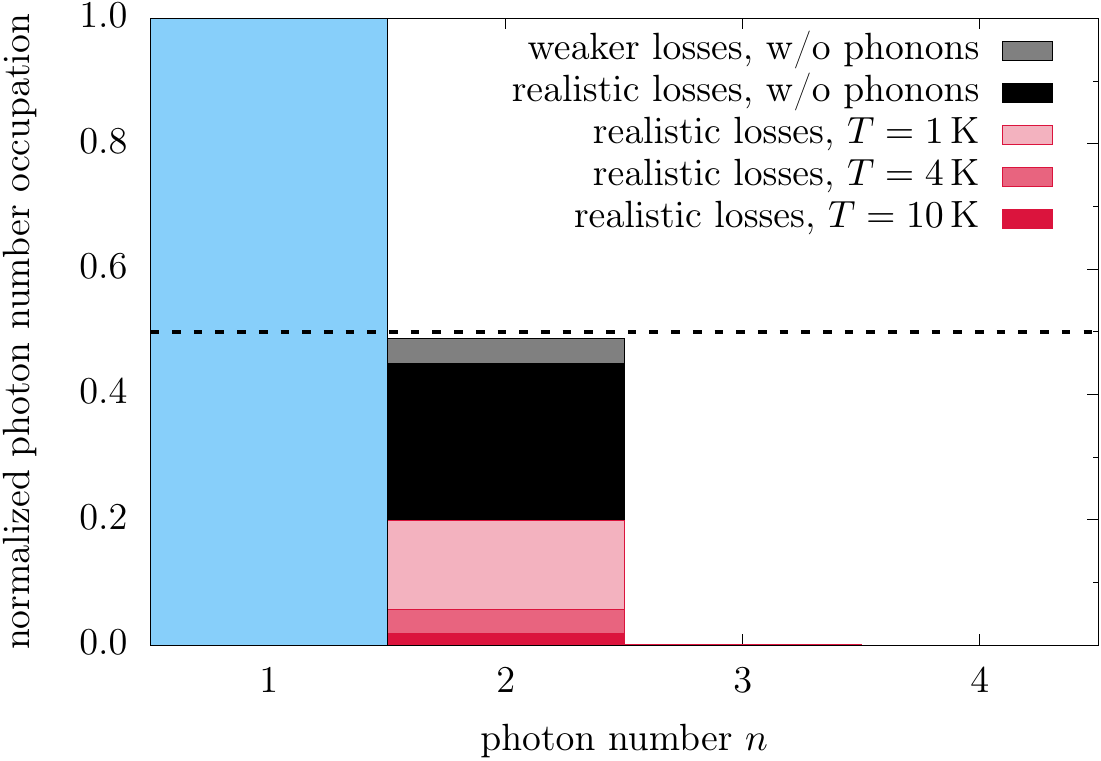}
\caption{The stationary photon number occupation normalized to its value at $n=1$ for the QD--cavity system.
While the data labeled 'realistic losses' is obtained using the parameters listed in Sec.~\ref{subsec:phonons}, weaker losses of $\gamma=0.01g$ and $\kappa=0.1g$ were chosen following Ref.~\cite{Munoz2014} for the calculation shown in gray.
Note that in the phonon-free case, the absolute values of the Fock state with $n=1$ are $0.016$ for the weaker losses and $0.003$ for the realistic parameter set.}
	\label{fig:ph_distr_QDC}
\end{figure}

\begin{figure}[t]
	\centering
	\includegraphics[width=\columnwidth]{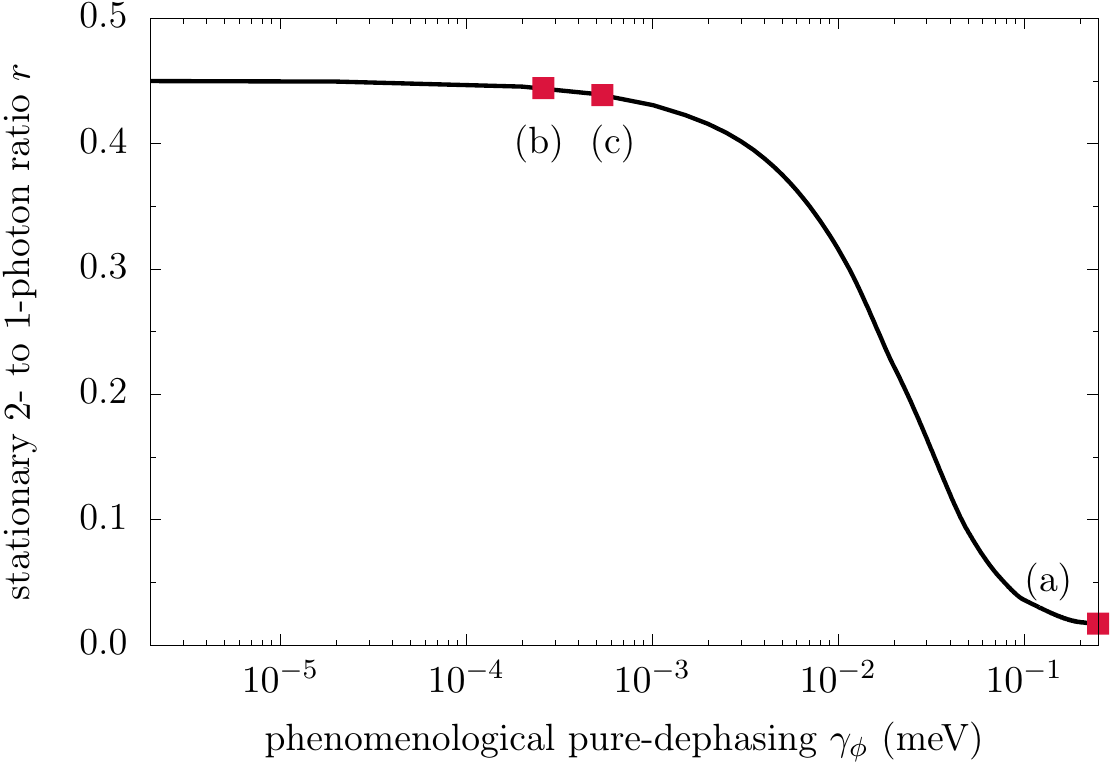}
\caption{The stationary ratio between the 2- and the 1-photon occupation in the QD--cavity system with the phonon influence approximated by a Lindblad operator with a phenomenological pure dephasing rate $\g_\phi$ instead of the microscopic Hamiltonian $H_{\t{Ph}}$ in Eq.~\eqref{eq:H_Ph}, cf. main text.
(a) $\gamma_\phi$ corresponding to the full driven Jaynes--Cummings model at $T=4\,$K.
(b) $\gamma_\phi$ corresponding to a Jaynes--Cummings dynamics with $n=1$. 
(c) $\gamma_\phi$ corresponding to a Jaynes--Cummings dynamics with $n=2$.}
	\label{fig:pd_ratio}
\end{figure}

The analysis in terms of laser-dressed states reveals the fact that the $2$-photon bundle resonance at $\hbar\D\w_{\t{LX}}=-0.51\,$meV originates from a $2$-photon process \cite{Munoz2014}, in this case a transition from $|-,0\>$ to $|+,2\>$.
The study of the influence of the phonons on this resonance shows that already at $4\,$K [inset of Fig.~\ref{fig:stat_occ_QDC_triple}(b)], it is strongly suppressed.
The occupation of $n=1$ strongly rises around the resonance peak.
Apparently, phonon-assisted $1$-photon transitions are favored against the $2$-photon process creating the bundle.
In particular, the characteristic $1/n$ fingerprint [cf., Eq.~\eqref{eq:1/n}] of the number distribution is violated.

To illustrate this point in more detail, the stationary photon number distribution normalized to its value at $n=1$ is shown in Fig.~\ref{fig:ph_distr_QDC}.
First of all, it is interesting to note that the ideal bundle statistics $\propto 1/n$ is only observed for loss parameters weaker than the realistic, state-of-the-art values (cf. gray data in Fig.~\ref{fig:ph_distr_QDC}).
This parameter set consists of $\gamma=0.01g=0.3\,$ns$^{-1}$ and $\kappa=0.1g=3\,$ns$^{-1}$, following the values chosen in Ref.~\cite{Munoz2014}.
Already the slightly higher values chosen in our work in accordance with current experiments (cf. Sec.~\ref{subsec:phonons}) lead to a ratio of the stationary $2$- to the $1$-photon occupation
\begin{align}
r:=\lim_{t\to\infty}\frac{\<|2\>\<2|\>(t)}{\<|1\>\<1|\>(t)}
\end{align}
equal to $0.45$.
Thus, the ratio deviates from the target of $0.50$, which is a necessary indicator for an $N=2$ bundle.
The phonon coupling pushes this value down to $r=0.20$ already at $T=1\,$K.
For higher temperatures up to $10\,$K, $r$ swiftly approaches zero and the $2$-photon bundle fingerprint cannot be observed anymore.
The $N$-photon bundle statistics ($1<N<\infty$) therefore seems to be hard to find in state-of-the-art QD--cavity systems.

Our finding that due to the phonon influence the occupation of the $n=2$ Fock state compared with the $n=1$ state is much lower than expected for an $N=2$ bundle does, however, not mean that $2$-photon emission features are precluded from observation.
The latter can still be made prominent, e.g., by spectrally filtering the emission as has been shown in Ref.~\onlinecite{Munoz2018}.

\subsection{Comparison with a phenomenological dephasing model}
\label{subsec:gamma_phi}

The phonon environment has a drastic influence on the $N$-photon bundle statistics as shown in the previous section for the case $N=2$.
Already at a low temperature of $T=1\,$K the $1/n$-distribution characteristic for the bundle [cf., Eq.~\eqref{eq:1/n}] is not recognizable anymore (cf., Fig.~\ref{fig:ph_distr_QDC}).
This result was obtained within a microscopic model of the phonon influence.
In contrast, in Ref.~\cite{Munoz2014}, the dephasing has been analyzed using a phenomenological Lindblad operator $\L_{\vert X\rangle\!\langle X\vert,\gamma_\phi}$.

It is therefore instructive to compare the microscopic model with the phenomenological one to check whether the latter is valid.
On first sight, we find a quite different behavior:
for the phenomenological model taking values for the corresponding Lindblad rate $\gamma_\phi$ from the literature on semiconductor QD--cavity systems, the impact of pure dephasing is almost negligible \cite{Munoz2014}.

To analyze this in more detail, we have plotted results of the phenomenological model in Fig.~\ref{fig:pd_ratio}, which shows the stationary ratio $r$ as a function of the phenomenological pure-dephasing rate $\gamma_\phi$ which is incorporated into the model by the addition of the Lindblad operator $\L_{\vert X\rangle\!\langle X\vert,\gamma_\phi}$ instead of the microscopic Hamiltonian model $H_{\t{Ph}}$.
Indeed, in that approximation a large plateau range is found where the ratio stays essentially at its phonon-free value of $r=0.45$ (cf., also Fig.~\ref{fig:ph_distr_QDC}).

To assess, what $\gamma_\phi$ should be chosen in the reduced model to best approximate the full phonon effect, we apply the following procedure:
We compare the exciton dynamics resulting from the full calculation (where phonons are included by $H_{\t{Ph}}$) with the phenomenological model (where $H_{\t{Ph}}$ is replaced by $\L_{\vert X\rangle\!\langle X\vert,\gamma_\phi}$) and vary $\gamma_\phi$ until the envelopes of the two dynamical results essentially match.
Note that we set $\kappa=\gamma=0$ for this procedure to extract the pure phonon influence on the dynamics.
Furthermore, this comparison is conducted for the all-resonant case, i.e., $\Delta\w_{\t{LX}}=\Delta\w_{\t{CX}}=0$.
We perform this procedure at $T=4\,$K for three different cases and mark the resulting rate $\gamma_\phi$ by red squares in Fig.~\ref{fig:pd_ratio}:
(a) Driven Jaynes--Cummings system with the initial state $|G,0\>$, resembling the closest approximation to the full calculation, (b) Jaynes--Cummings system without driving ($f=0$) for the initial state $|G,1\>$, and (c) same as (b) but with $|G,2\>$ as the initial state.
The three extracted rates (cf., Fig.~\ref{fig:pd_ratio}) indicate that a very large pure-dephasing rate of the order of $10^{-1}\,$meV is necessary to reproduce the dynamics of the full microscopic model [cf., red square labeled with (a)].
With such a large rate, the ratio $r$ is close to zero, meaning that no $2$-photon bundle statistics is observed in accordance with the results of the full model at $T=4\,$K (cf., Fig.~\ref{fig:ph_distr_QDC}).

The reason for such a significant increase in $\gamma_\phi$ lies in the impact of the pure dephasing mechanism, which gains in strength for larger Rabi frequencies related to the effective couplings present in the system.
While in (b) and (c) the cavity Rabi frequency amounts to $2g\sqrt{n+1}$ with $n$ the number of photons present in the cavity, the driving $f\gg g$ introduces the highest transition frequency in (a).
In Fig.~\ref{fig:pd_ratio}, it becomes clear that the pure-dephasing rate increases with larger effective coupling, in accordance with earlier observations in the case of a microscopic description of phonons \cite{Machnikowski2004,Vagov2007,Glaessl2012a}.
The values of $\gamma_\phi$ in (b) and (c) are of the order of experimentally found pure-dephasing rates for strong QD--cavity coupling like the one studied here (cf., Sec.~\ref{subsec:phonons}), but without external driving.
Choosing such values for the rate indeed results in a marginal influence of pure dephasing, since both points lie well inside the plateau region.

Thus, the conclusion in Ref.~\cite{Munoz2014} that dephasing does not significantly affect the $N$-photon bundle generation can be traced back to the fact that values for dephasing rates have been considered that are no longer applicable in the regime of very strong driving as required for this protocol.
The physical reason lies in the fact that an optically driven system is influenced by the phonon Hamiltonian in a profoundly different way than its non-driven counterpart:
while phonons cannot induce transitions between the two electronic states in the undriven case, they can lead to transitions between the laser-dressed states, which are the eigenstates of the driven two-level system.
In essence, the dephasing rate depends on the driving strength.
A quadratic dependence $\gamma_\phi\propto f^2$ can be derived in a weak-coupling limit \cite{Nazir2016}.

In conclusion, a phenomenological pure dephasing model is also able to qualitatively predict that the characteristic statistical fingerprint of $N$-photon bundles is violated.
The challenge is the choice of a proper rate, which has to be calibrated to the full phonon system.

\section{Results: Superconducting qubit--microwave resonator systems}
\label{sec:superconducting}

Superconducting qubit--microwave resonator systems have been successfully used to demonstrate the on-demand preparation of various highly nonclassical photon states, such as Fock states \cite{Hofheinz2008}, superpositions thereof, and Voodoo cat states, i.e., coherent superpositions of three coherent states \cite{Hofheinz2009}.
In none of these experiments, a significant impact of pure dephasing was reported.

\begin{figure}[t!]
	\centering
	\includegraphics[width=\columnwidth]{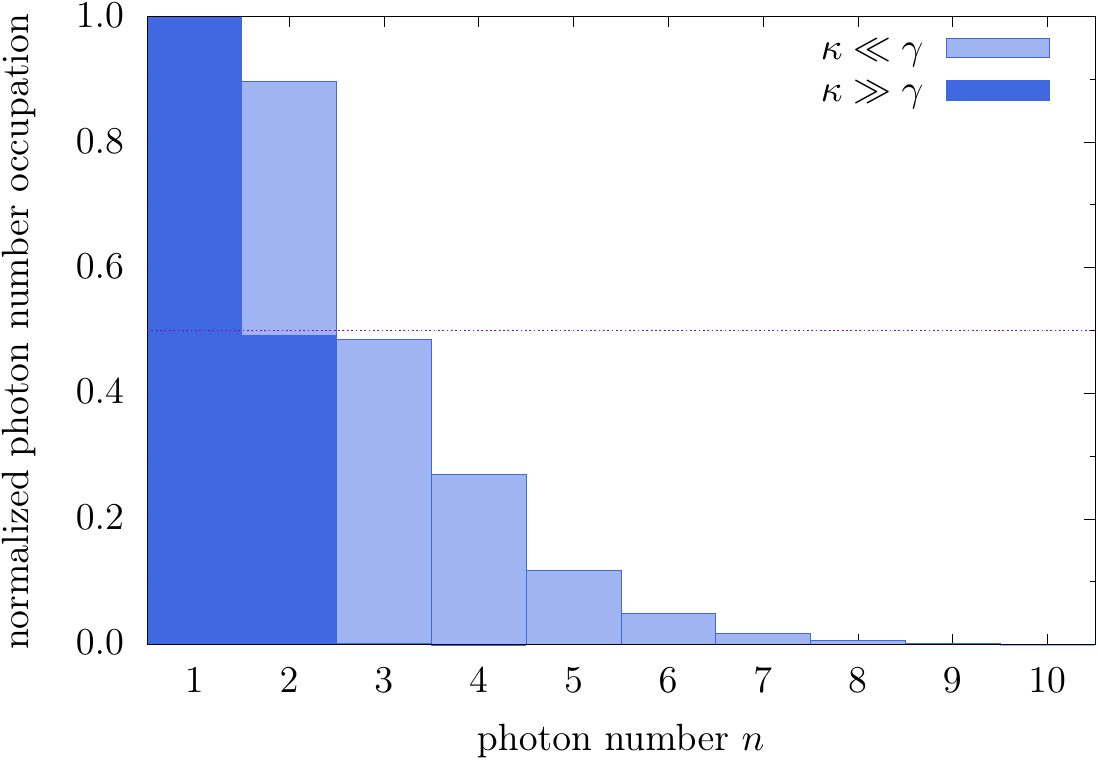}
\caption{The stationary photon number occupation normalized to its value at $n=1$ for the superconducting qubit--microwave system.
The data labeled $\kappa\ll\gamma$ is obtained using the parameters from Sec.~\ref{subsec:SC}.
In dark blue, the result of a calculation with a cavity loss rate two orders of magnitude larger than in Sec.~\ref{subsec:SC} is shown, namely $\kappa=0.1g=7.76\gamma$, cf. Fig.~\ref{fig:ph_distr_QDC}.}
	\label{fig:ph_distr_superconducting}
\end{figure}

For state-of-the-art superconducting systems \cite{Hofheinz2009}, the resonator losses are much smaller than the decay of the qubit ($\kappa\ll\gamma$ as in Sec.~\ref{subsec:SC}).
Again, the $2$-photon bundle resonance is achieved by an external excitation tuned according to Eq.~\eqref{eq:bundle_resonance}.
The resulting photon number distribution is shown in Fig.~\ref{fig:ph_distr_superconducting}, normalized to its value at $n=1$ (light blue bars).
Surprisingly, no bundle statistics is found, as the photons are able to climb up the Jaynes--Cummings ladder instead.
In particular, the characteristic cutoff for $n>N=2$ is not observed.
The reason lies in the fact that the radiative decay $\gamma$ can induce transitions from $|+,n\>$ to $|-,n\>$.
From the latter, additional photons can be emitted to break the cutoff and reach higher $n$.

The failure of the superconducting qubit to show the statistical fingerprint can be traced back to the lack of resonator losses $\kappa$ in comparison to radiative decay $\gamma$.
Indeed, if we consider a resonator loss rate much larger (following Ref.~\cite{Munoz2014}, $\kappa=0.1g$ has been chosen, cf., also Fig.~\ref{fig:ph_distr_QDC} for this specific choice), we can obtain a near-perfect $2$-photon bundle statistics.
The resulting photon number distribution (cf., dark blue bars in Fig.~\ref{fig:ph_distr_superconducting}) indeed shows a near-perfect $2$-photon bundle fingerprint, with $r=0.49$ and no occupation for $n>2$.
This means that though much effort is usually invested into resonators of better quality, here the use of a bad resonator is mandatory.

\begin{figure}[t!]
	\centering
	\includegraphics[width=\columnwidth]{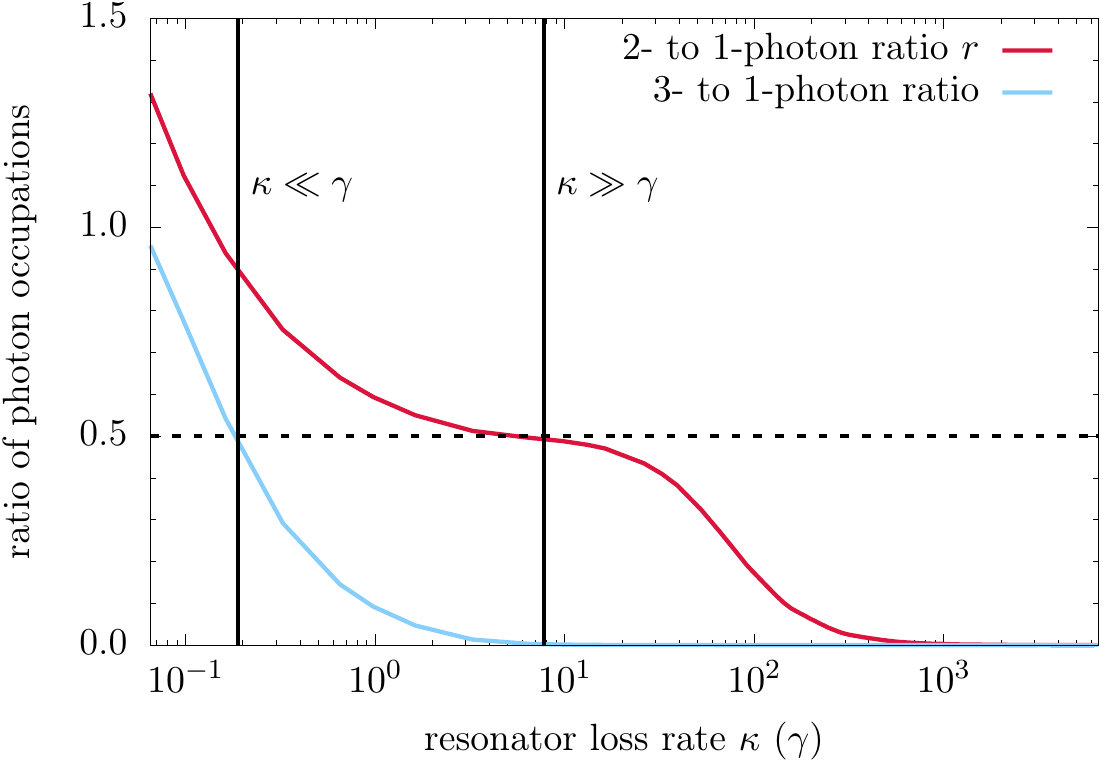}
\caption{The stationary ratios of the $2$- to $1$-photon occupations $r$ and of the $3$- to $1$-photon occupations as functions of the resonator loss rate $\kappa$ (in units of $\gamma$) for the superconducting qubit--microwave system.
The two vertical black lines mark those values of $\kappa$, which are used to obtain the corresponding data in Fig.~\ref{fig:ph_distr_superconducting}.
The dotted black line shows the target value of $0.5$ for the ratio $r$.}
	\label{fig:kappa_ratio}
\end{figure}

To analyze the impact of the losses in more detail, we study the bundle statistics as a function of the resonator losses.
To this end, the $2$- to $1$-photon ratio $r$ is shown as a function of $\kappa$ in Fig.~\ref{fig:kappa_ratio} as well as the $3$- to $1$-photon ratio, which should vanish for an ideal $2$-photon bundle emission due to the cutoff for $n>N=2$.
Indeed, these two quantities confirm that the chosen value of $\kappa=0.1g=7.76\gamma$ lies well within a plateau region of $r\approx 0.5$ and a vanishing occupation for $n>2$.
While resonator losses too low compared to the decay of the qubit results in the occupation of states with $n>2$, using very low-quality resonators with $\kappa\gtrsim20\gamma$ (cf., Fig.~\ref{fig:kappa_ratio}) leads to a drastic reduction of $r$ and thus a statistics, which does not show the bundle fingerprint anymore.
While constructing resonators of better quality is always an experimental challenge, creating a bad resonator should be a lesser problem.
Thus, superconducting qubit--microwave resonator systems are indeed suitable candidates for sources of $N$-photon bundles, in agreement with Ref.~\onlinecite{Ma2021}.

\section{Conclusion}
\label{sec:Conclusion}

We have studied the $N$-photon bundle statistics in two solid-state platforms:
semiconductor quantum-dot--cavity systems and superconducting qubit--microwave resonator systems.

In quantum-dot--cavity systems, pure dephasing is induced by longitudinal acoustic phonons.
We have found that even at low operating temperatures of a few kelvin, the characteristic bundle statistics [cf., Eq.~\eqref{eq:1/n}] cannot prevail for $N=2$, thereby implying that a corresponding statistics for $N>2$ is also out of reach with current state-of-the-art samples.
The reason is the considered driving regime that is required to address the bundle resonance, which also favors the phonon activity in the electronic subsystem of the quantum dot.

In contrast, superconducting qubit--microwave resonator systems are suitable candidates for the observation of the $N$-photon bundle statistics.
Here, the pure dephasing does not play a notable role.
However, the quality of the resonator should not be too high to facilitate the emission of photon bundles.

\acknowledgments
This work was funded by the Deutsche Forschungsgemeinschaft (DFG, German Research Foundation) -- project No. 419036043.

\appendix
\section{Resonance peaks for $N\to\infty$ and $N=1$}
\label{app:limits}

\begin{figure}[t]
	\centering
	\includegraphics[width=\columnwidth]{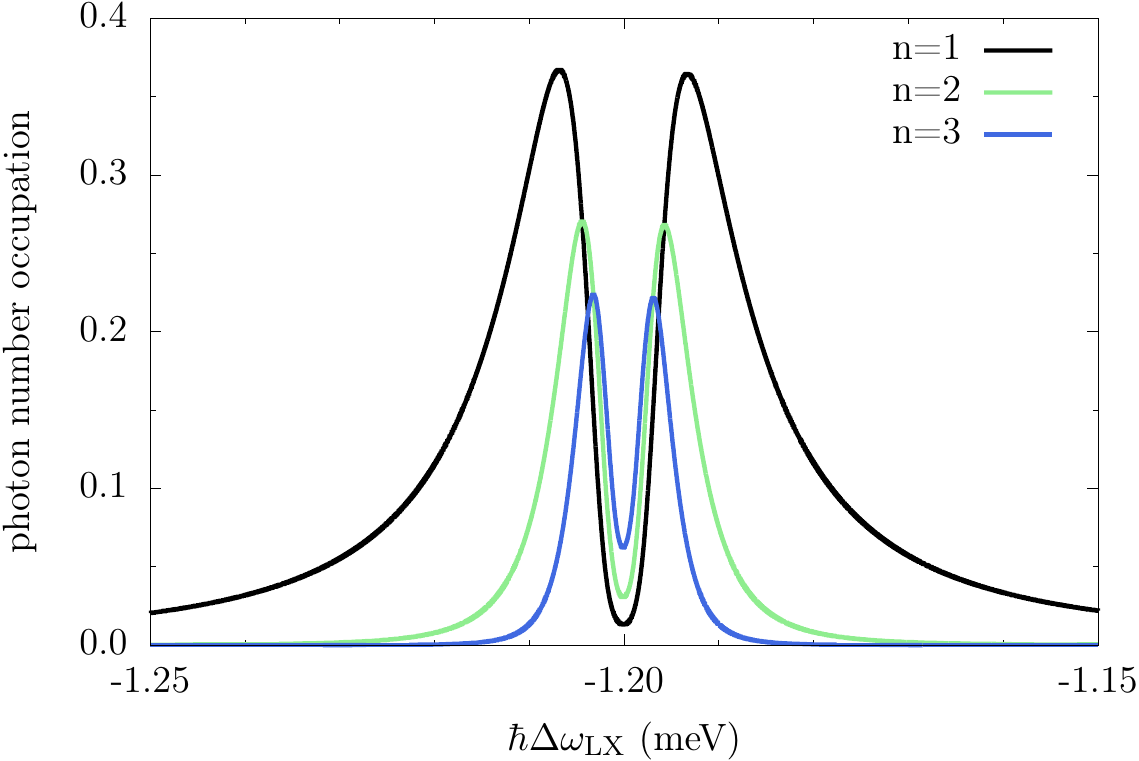}
\caption{Stationary photon number occupation in the QD--cavity system as a function of the laser--exciton detuning $\D\w_{\t{LX}}$ without taking phonon effects into account.
This is a magnification of the resonance peak for $N\to\infty$ in Fig.~\ref{fig:stat_occ_QDC_triple}(a).
On this scale, the double-peak structure and the reversal of the photon order at its center are well visible.
}
	\label{fig:magnified_N_infty}
\end{figure}

Since the peaks at $\hbar\D\w_{\t{LX}}=-1.2\,$meV and $0.08\,$meV are the most striking features in Fig.~\ref{fig:stat_occ_QDC_triple}, we shall discuss them in some detail in this appendix.
This will give additional insights into the physics taking place in this parameter regime in general, although the analysis reveals that these peaks are not related to the bundles which are the main target of our paper.
The most prominent peak in Fig.~\ref{fig:stat_occ_QDC_triple} at $\hbar\D\w_{\t{LX}}=\hbar\D\w_{\t{CX}}=-1.2\,$meV is obtained in the limit $N\to\infty$ and corresponds to a process where the photon energy in a frame rotating with the laser frequency is $\hbar\D\w_{\t{CL}}=0$ and the system can climb up the photon ladder from $|-,n\>$ to $|-,n+1\>$, such that a Poissonian distribution with respect to $n$ emerges.
Note that one observes a double-peaked structure at this resonance in Fig.~\ref{fig:stat_occ_QDC_triple}.
At its center, the order of the photon occupations is reversed, i.e., the occupation of $n=2$ is higher than that of $n=1$, consistent with a Poissonian with an average photon number of $\<n\>=6.6$ and a maximum occupation of $0.15$ at $n=6$.
A magnification of this peak, where the reversal of the photon order is visible, is replotted in Fig.~\ref{fig:magnified_N_infty}.
An analysis of the corresponding Wigner function \cite{Hofheinz2009,Cosacchi2021a} (not shown here) confirms that the corresponding state is a (Glauber) coherent state.

The peak at $\hbar\D\w_{\t{LX}}\approx0.08\,$meV corresponds to a $1$-photon bundle resonance, i.e., a $1$-photon Fock state, and also results from a one-photon process.
But in contrast to the previously discussed case, the photon is emitted by the transition from $|-,0\>$ to $|+,1\>$.
Due to an energy mismatch between the photon energy and the transition between $|+,1\>$ and $|\pm,2\>$, no further photons are put into the cavity, as can be seen in the stationary occupations of this peak in Fig.~\ref{fig:stat_occ_QDC_triple}(a).
This effect is commonly known as the photon blockade \cite{Birnbaum2005}.

The phonon influence on the occupations at $T=4\,$K as shown in Fig.~\ref{fig:stat_occ_QDC_triple}(b) could not be more different for these two resonances.
The first one for $N\to\infty$ at $\hbar\D\w_{\t{LX}}=-1.2\,$meV is hardly influenced by phonons at all.
Indeed, the photon number distribution remains Poissonian with a slightly lower average photon number of $\<n\>=5.6$ and a similar maximum occupation of $0.16$ at $n=5$.
The reason lies in the fact that the photons are emitted from transitions, where the electronic (laser-dressed) state remains $|-\>$ and does not change.
Since this is the energetically lower dressed state and at temperatures below a few tens of kelvins phonon absorption is highly unlikely, phonons have only a slight influence on the stationary photon distribution.

On the other hand, the second peak at $\hbar\D\w_{\t{LX}}\approx0.08\,$meV for $N=1$ experiences strong phonon-enhancement, since the photon blockade is spoiled.
The energy mismatch between $|+,n\>$ and $|-,n\>$ is now bridged by phonon emission, which is possible for all temperatures down to absolute zero, and a subsequent resonant transition to $|+,n+1\>$ can take place.
Therefore, the phonon coupling drives the occupation of higher-order Fock states beyond $n=1$ \cite{Cygorek2017,Cosacchi2020a}.

\bibliographystyle{APPA}
\bibliography{bib}
\end{document}